\documentclass{aa}
\usepackage{graphicx}
\def \ha   {H$\alpha$}
\def \oiii {[O~{\sc iii}]5007~\AA}
\def \nii  {[N~{\sc ii}]6584~\AA}
\def \NII  {[N~{\sc ii}]~6548~\&~6584~\AA}

\newcommand{\NiII}{[Ni~{\sc ii}]~7378 \& 7412~\AA}
\def \tena#1 #2 {\ifmmode{#1 \times 10^{#2}}\else{$#1 \times 10^{#2}$}\fi}
\def \kms  {\ifmmode{~{\rm km\,s}^{-1}}\else{~km s$^{-1}$}\fi}
\def \vhel {\ifmmode{~V_{{\rm HEL}}}\else{~$V_{{\rm HEL}}$}\fi}
\def \vlsr {\ifmmode{~V_{{\rm LSR}}}\else{~$V_{{\rm LSR}}$}\fi}
\def \vsys {\ifmmode{~V_{{\rm sys}}}\else{~$V_{{\rm sys}}$}\fi}
\def \vexp {\ifmmode{~V_{{\rm exp}}}\else{~$V_{{\rm exp}}$}\fi}

\def \deg  {\ifmmode{^{\circ}}\else{$^{\circ}$}\fi} 
\def \msun {\ifmmode{{\rm\ M}_\odot}\else{${\rm\ M}_\odot$}\fi}
\def \myr  {\ifmmode{{\rm\ M}_\odot{\rm\ yr}^{-1}}
         \else{${\rm\ M}_\odot$ yr$^{-1}$}\fi}
\def\mnras{MNRAS}
\def\apj{ApJ}
\def\apjs{ApJS}
\def\aap{A\&A}

\def\pasp{PASP}

\begin{document}

\title{Candidates for giant lobes projecting from the LBV stars P~Cygni
and R~143}

\author{J. Meaburn\inst{1} 
     \and   P. Boumis\inst{2}
     \and M. P. Redman\inst{3}
     \and   J. A. L{\'o}pez\inst{4}  
     \and   F. Mavromatakis\inst{5}
          }

\offprints{Prof. J. Meaburn}
\authorrunning{J. Meaburn et al.}
\titlerunning{Giant lobes of LBV stars.}
\institute{Jodrell Bank Observatory, University of Manchester, 
Macclesfield SK11 9DL, UK.
\and Institute of Astronomy \& Astrophysics, National Observatory of
Athens, I. Metaxa \& V. Paulou, P. Penteli, GR-15236 Athens, Greece
\and Dublin Institute for Advanced Studies, School of Cosmic Physics, 
5 Merrion Square, Dublin 2, Republic of Ireland.
\and Instituto de Astronomia, UNAM, Apdo. Postal 877. Ensenada, B.C. 22800,
M\'{e}xico.
\and University of Crete, Physics Department, P.O. Box 2208, 710 03 
Heraklion, Crete, Greece.
}

   \date{Received ; accepted }

\abstract{Deep, wide--field, continuum--subtracted, images in the light of the
\ha\ + \NII\ and \oiii\ nebular emission lines have been obtained of the
environment of the Luminous Blue Variable (LBV) star P~Cygni.
A previously discovered, receding, nebulous filament along PA~50\deg\ 
has now been
shown to extend up to 12\arcmin\ from this star. Furthermore, in the
light of \oiii, a southern counterpart is discovered as well as
irregular filaments on the opposite side of P~Cygni.

Line profiles from this nebulous complex indicate that this 
extended nebulosity is similar to that associated with
middle--aged supernova remnants. However, there are several indications
that it has originated in P~Cygni and is not just a chance superposition
along the same sight--line. This possibility is explored here and comparison
is made with a new image of the LBV star R~143 in the LMC from which similar
filaments appear to project.

The dynamical age of the P~Cygni giant lobe of $\approx$ 5$\times$10$^{4}$ yr
is consistent with both the predicted and observed durations of
the LBV phases of 50\msun\ stars after they have left the main sequence. 
Its irregular shape may have been determined by the cavity
formed in the ambient gas by the energetic wind of the star, and
shaped by a dense torus,
when on the main sequence. 

The proper motion and radial velocity of P~Cygni, 
with respect to its local environment,
could explain the observed angular and kinematical shifts of the star
compared with the giant lobe.

\keywords{CSM:general--CSM: LBV stars --CSM: individual objects:
 P~Cygni--R~143}}

\maketitle

\section{Introduction}

The circumstellar environment of the proto-typical Luminous Blue
Variable star (LBV - Conti 1984; Humphreys 1989; Davidson, Moffat \&
Lamers 1989) P~Cygni has been revealed at optical wavelengths in the
work presented in six recent papers (Johnson et al. 1992; Barlow et
al 1994; Meaburn et al 1996; O'Connor, Meaburn \& Bryce 1998; Meaburn,
L\'{o}pez \& O'Connor 1999; Meaburn et al 2000). 
Two nearly spherical, but complex,
circumstellar shells were discovered. The bright \NII\ and \NiII\
emitting, 22\arcsec\ diameter inner shell (IS) was found (Barlow et al
1994) to be surrounded by a fainter \NII\ emitting, 1.6\arcmin\
diameter, outer shell (OS). The dynamical ages of the IS (Barlow et
al 1994) and the OS (Meaburn et al 2000), for a distance to P~Cygni of
1.8~kpc (van Schewick 1968; Lamers, de Groot \& Cassatella 1983) were
derived from their expansion velocities as 880 and 2400 yr
respectively.  This would place the creation of both of these
shells as well before the great outburst of 1600 {\sc ad} (de Groot 1969).
Humphreys and Davidson (1994) emphasise that knowledge of P~Cygni's
eruptive `geyser--like' behaviour prior to this date is unknown.

Potentially as interesting, is the presence of a filamentary, line
emitting,
giant lobe (GL), discovered by O'Connor, Meaburn \& Bryce (1998)
which could be a relic of the activity of P~Cygni close, or even prior,
to
its entry into its LBV phase.  
In the later work by Meaburn et al (2000) the  
northern ridge of GL had been traced for 7\arcmin, along PA $\approx$
50\deg, from P~Cygni and shown to connect morphologically and 
kinematically
with the receding side of the  OS. However,  the previous observations
in Meaburn et al (1999 \& 2000)
of the kinematical behaviour 
of this northern
ridge of GL
strengthen, but do not absolutely confirm, the suggestion that it
is directly associated with P~Cygni (Meaburn et al
1999) and not a chance superposition along the same sight-line.  It
is also significant that a morphologically similar feature (Meaburn 2001) 
to the P~Cygni GL
has since been found (see Smith et al 1998 - their figure 2) to be
apparently projecting from the LBV star R143 in the Large Magellanic Cloud.

The original observations by O'Connor et al (1998) 
of the P~Cygni GL, revealing its north eastern
filamentary arc, were made in the light of the \nii\ 
nebular emission line. In the present paper 
these have  been supplemented by deep, wide--field,
continuum--subtracted images in the light of \ha\ + \NII\ and more
significantly \oiii. Although many new, but connected, features 
are reported in the present
paper the designation GL will be used to describe the whole of
this extended nebular complex that has now been revealed around P~Cygni. 
Previously obtained
line profiles will also be compared with these most recent images.
Similarly, a new deep image of the candidate GL surrounding R~143
in the LMC will be presented and compared to the motions 
measured by Weis (2003).

\section{Observations and results}

The new \ha\ plus \NII,  \oiii\ and continuum images of P~Cygni were 
obtained on 25 June and 1 July
2003 with the 89\arcmin\ $\times$ 89\arcmin\ field (5\arcsec\ per
pixel) 0.3 m Schmidt Cassegrain telescope at the Skinakas Observatory,
Crete, Greece of the nebulosity surrounding P~Cygni.  
The integration times were 4800 s (for \oiii) and 7200 s (for \ha + \NII). 
The corresponding continuum images 
were subtracted
from those containing the emission lines to eliminate the confusing
star field (see Mavromatakis et al 2002 for details of this technique).

For the first time the extent of the \oiii\ emission from the giant lobe
surrounding P~Cygni is revealed. This is shown as contours with 
linear intervals in Fig. 1 
compared with the grey-scale representation of the \ha\ plus \NII\ emission.
The bright (V = 4.8) P~Cygni created strong ghost images in all of these
exposures  causing  confusion in the areas blanked out in Fig. 1.
The astrometric solutions were calculated for each
field individually using reference stars from the Hubble Space
Telescope (HST) Guide Star Catalogue (Lasker et al 1999). 

The
spectrophotometric standard stars HR5501, HR7596, HR7950, and HR8634
(Hamuy et al 1992 \& 1994) were used for absolute flux
calibration of the \oiii\ emission shown in Fig. 1. 
The [O~{\sc iii}] 5007~\AA\ fluxes measured in
different parts of the eastern and western complexes of the P~Cygni GL
are listed in Table~1.

The previously known (Meaburn et al 2000) 
7\arcmin\ long northern arc of the GL to the east 
of P~Cyg
has been shown in Fig.~1 in the \oiii\ and \ha\ + \NII\ emission lines 
to extend to 12\arcmin\ from the star. 
For the first time a southern counterpart to this northern arc
of the GL, as well as a complex western extension, can also be seen in Fig.~1.
The contrast is enhanced for the detection of this high 
excitation arc in the \oiii\ line 
against the confusing line emission phenomena in the lower excitation   
galactic background along the same sight--line.

The northern and southern ridges of the GL to the east of P~Cygni
can be seen in the position--velocity (pv) array of \nii\ profiles in Fig. 2, 
along slit length
A in Fig. 1, 
to have the characteristics of receding
walls of a common cavity. The western complex of the GL can also
be seen in the pv array of \nii\ profiles in Fig. 2, close to 
P~Cygni along slit B to be 
receding from the star with a radial velocity equal to that
of the far--side of the OS. The pv arrays in Fig. 2 are from 
the previous, more extensive,  spectral 
observations obtained with the Manchester Echelle Spectrometer 
on the 2.1--m San Pedro Martir telescope (Meaburn et al 1984 
\& 2003): these spectral observations are presented in full
in Meaburn et al (1999 \& 2000). The two pv arrays have been
selected from this larger sample  
for they permit direct comparison with the newly discovered regions of the
P~Cygni GL shown in Fig. 1. For instance the line profiles
of the GL south ridge and for the GL west complex can be appreciated
graphically  
in the pv arrays in Fig. 2 
for slits A and B repectively.

\begin{table}
\caption[]{Fluxes of the \oiii\ emission for  the brightest areas of the
P~Cygni GL.}
\label{fluxes}
\begin{flushleft}
\begin{tabular}{lllllll}
\hline
\noalign{\smallskip}
\bf Eastern complex & 3$\arcmin$ & 4$\arcmin$ & 6$\arcmin$ & 8$\arcmin$ &
10$\arcmin$ & 12$\arcmin$ \\
\hline
North  & 2.25 & 2.71 & 2.01 & 2.23 & 2.10 & 2.17 \\
South  &  & 1.91 &  & 1.85 &  & \\
\hline
\hline
\bf Western complex & 3$\arcmin$ & 3$\arcmin$.5 & 4$\arcmin$ & 4$\arcmin$.5
& 6$\arcmin$.5 &  \\
\hline
North & 2.41 & 2.17 & 2.73 & & & \\
North--west & 2.73 & &  & 1.95 & 3.76 \\
South--west & & & 2.30 & & \\
\hline
\end{tabular}
\end{flushleft}
${\rm }$ All arcmin values are away from P Cygni. \\
${\rm }$ Fluxes in units of $10^{-16}$ erg s$^{-1}$ cm$^{-2}$
arcsec$^{-2}$. \\
$^{\rm}$ Median values over a 55\arcsec $\times$ 55\arcsec\ box. \\
$^{\rm}$ Fluxes are uncorrected for interstellar extinction.
\end{table}

The image of the environment of the 
LBV star  R~143 in Figs. 3a \& b, to be compared with
that of P~Cygni in Fig. 1, 
was taken with the New Technology Telescope
(La Silla) through a 40 \AA\ bandwidth filter centred on 
\ha. The integration time was 60 s. Filaments of emission line nebulosity
shown in the deep presentation in Fig. 3a
apparently extend from  the star and connect with a bright ridge
of nebulosity in its close vicinity (Fig. 3b). The pixel size 
is equivelent to 0.34\arcsec\ x 0.34\arcsec.

\begin{figure*}
\centering
\includegraphics[height=0.4\textheight]{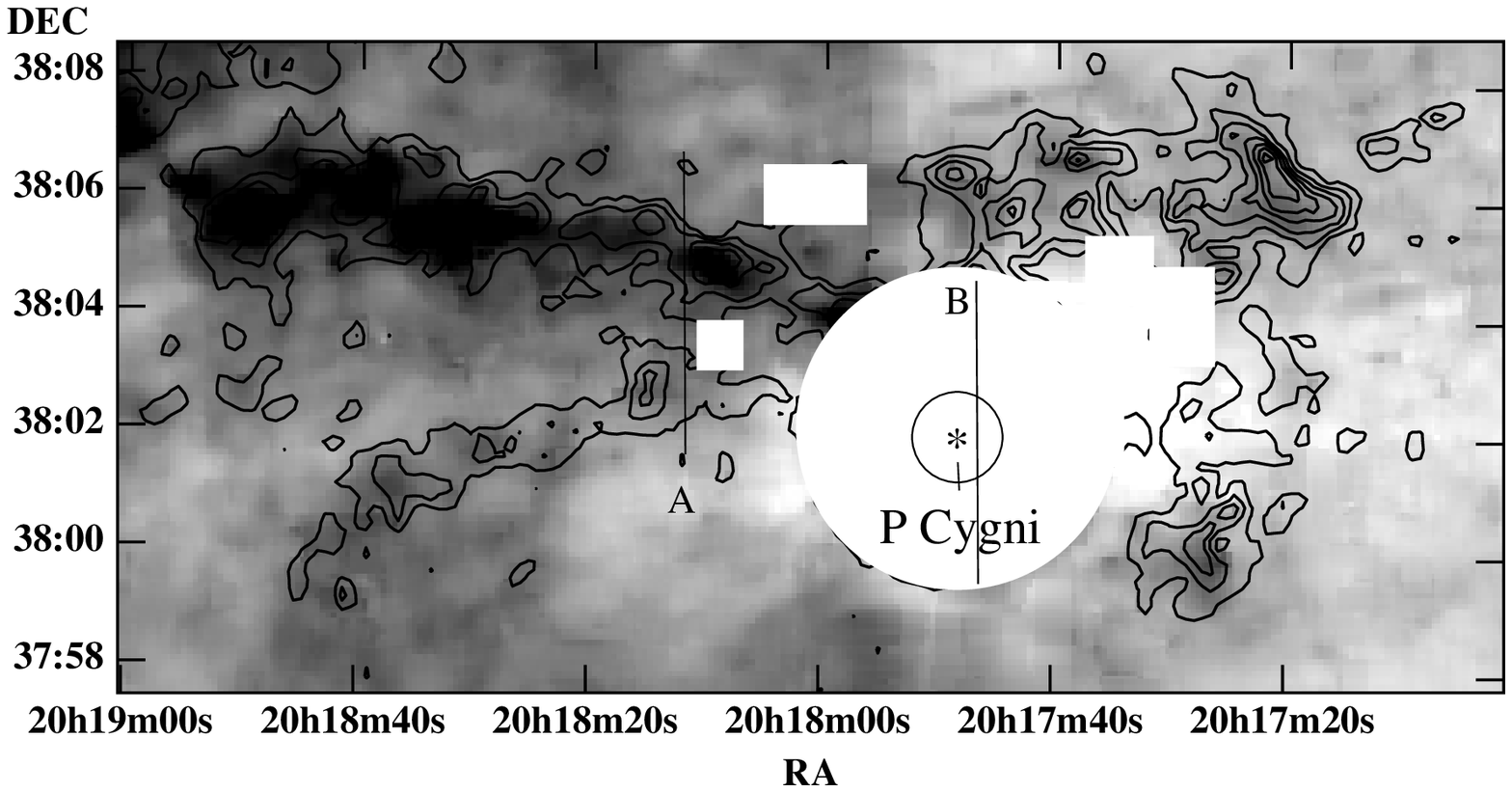}
\caption{The contours with linear intervals  are for the \oiii\ 
emission from the giant lobe around P~Cygni. These are
overlain on a negative grey-scale presentation of the \ha\ and \NII\
emission. Areas affected by the scattered light from the 
central star are blanked out. The black circle around P~Cygni
depicts the extent of the outer shell (OS) and the MES slit positions
A and B where the spectra in Fig. 2 were obtained are also shown. Coordinates 
are J2000.}
\label{Fig1} 
\end{figure*}

\begin{figure*}
%\centering
\includegraphics[height=0.6\textheight]{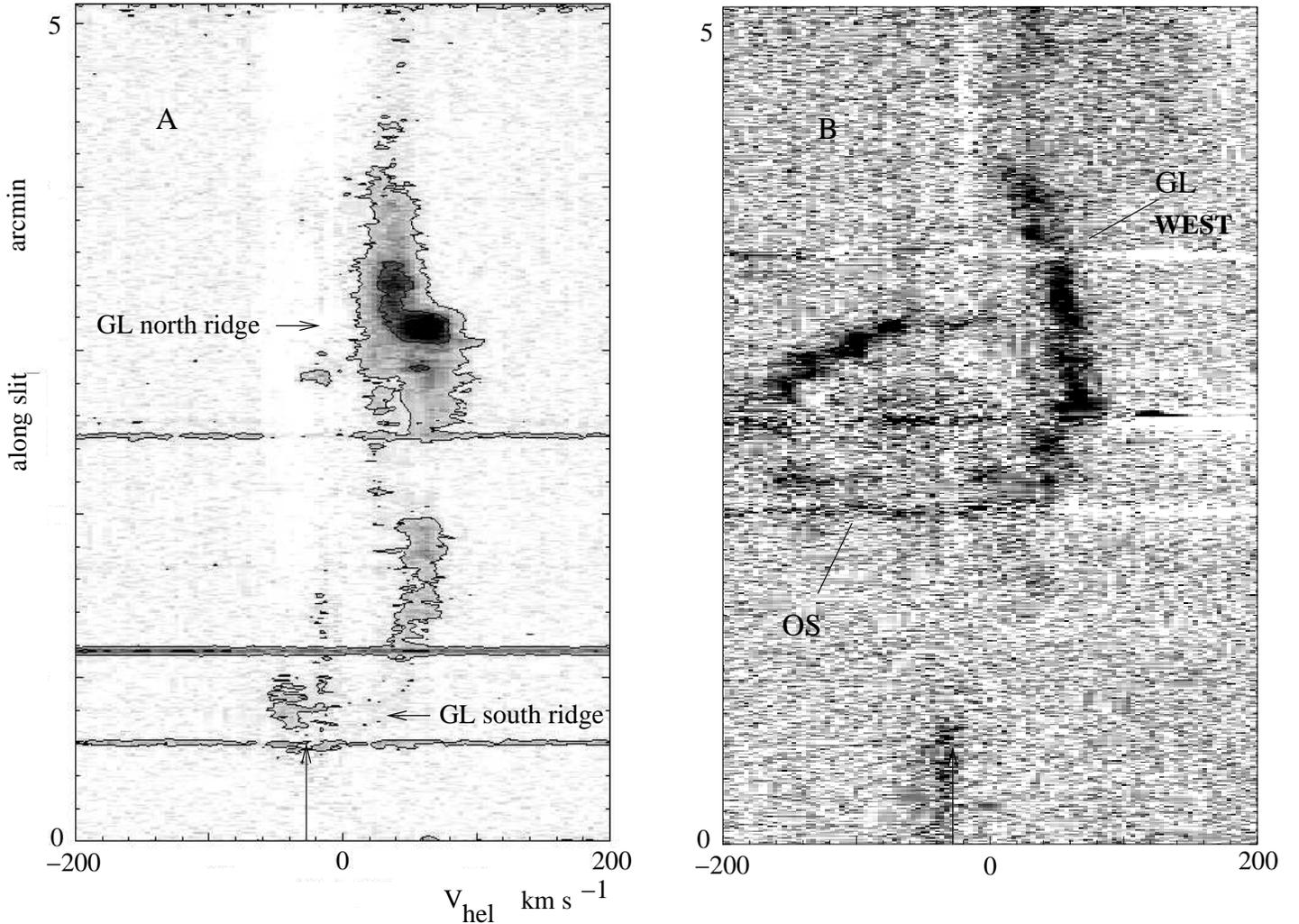}
\caption{Grey-scale representations of the position-velocity arrays
of \nii\ line profiles along the slit positions A and B (see Fig. 1)
are shown. The line profiles along slit length A over the northerly ridge 
(marked GL north) and the 
newly discovered southerly ridge (marked GL south) can be appreciated. 
The spectral features from the outer shell (marked OS) are distinguished 
from those of the westerly giant lobe (marked GL west) 
along slit length B. The
systemic heliocentric radial velocity (-26 \kms) of P~Cygni is arrowed.
This was given by the central velocity of the OS assuming
spherical expansion.
} 
\label{Fig2} 
\end{figure*}

\begin{figure*}
%\centering
\includegraphics[height=\textheight]{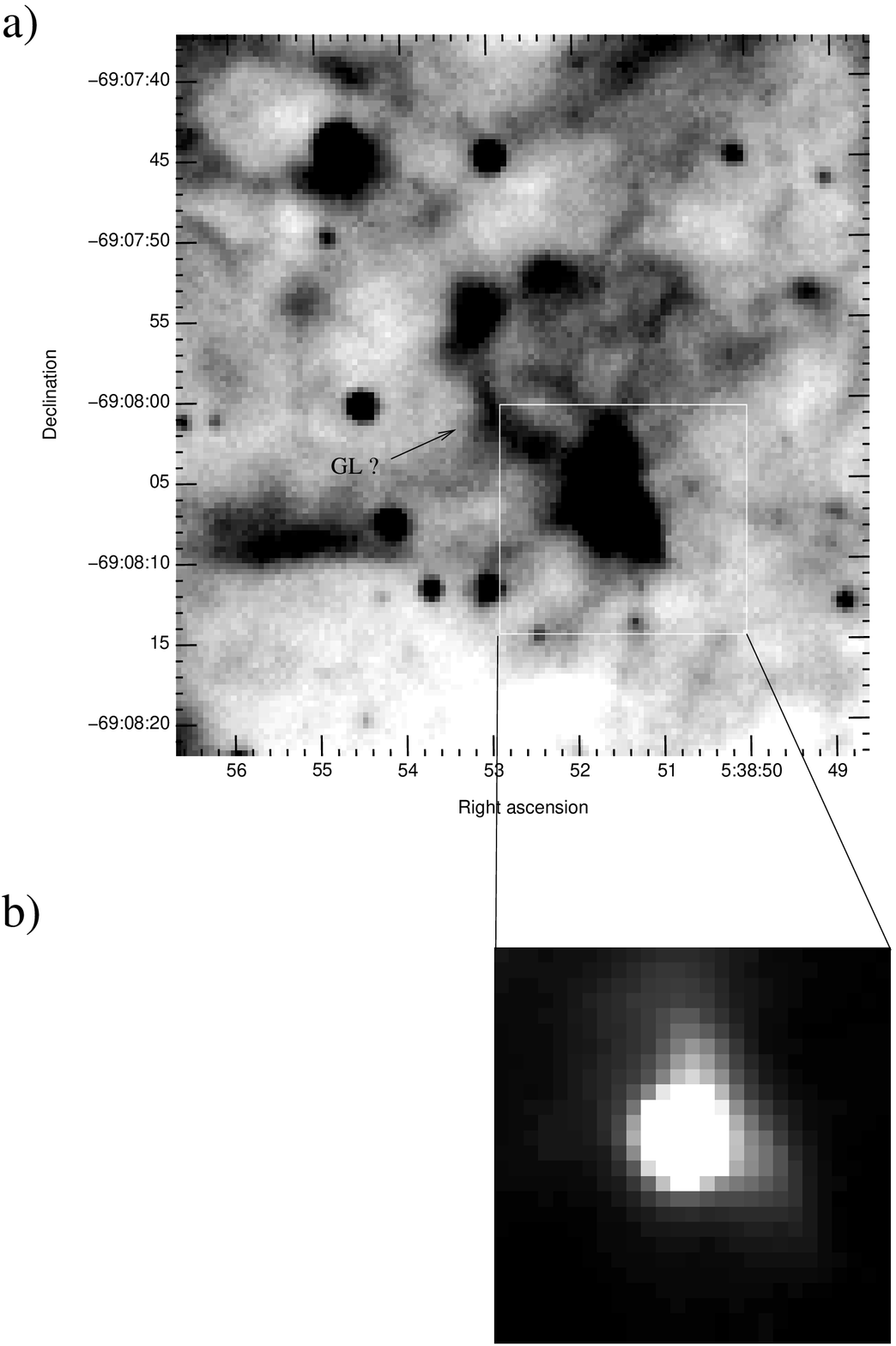}
\caption{a) The filaments that could possibly be giant lobes 
projecting from the LBV star R~143 in the LMC can be seen in this
\ha\ plus \NII\ image and are indicated by GL? b) These faint northern
filaments 
connect with the bright ridges near to this star. Coordinates are J2000.} 
\label{Fig3} 
\end{figure*}

\section{Discussion}

\subsection{Strength of association}

The primary consideration is 
still to evaluate the evidence that 
associates the apparent GLs around P~Cygni and R~143 with these stars;
as opposed to them being chance coincidences along the same sightlines.
Both stars are in crowded fields but overlapping filamentary 
structures (Figs. 1 \& 3)
 with similar dimensions
(6 sec~$\theta$ pc from P~Cygni and 4 sec~$\theta$ pc 
from R~143 where $\theta$ is the angle between the length of the GL and the 
plane of the sky) alone strengthens 
the possibility of 
associations; as does the discovery by Clark et al (2003) of the 
gaseous structure of the same size around the LBV candidate G24.73+0.69.  

The \nii\ line profiles (Meaburn et al, 1999 \& 2000) from the
northern arc of the eastern part of the proposed P~Cygni GL have
some of the characteristics expected of those from a collisionally
ionized filament of a middle aged supernova remnant; yet there is no
attendant non-thermal radio emission. In fact the radio emission
detected by Skinner et al (1997) appears to be of thermal origin 
(Meaburn et al 2000). In any case it would be a strangely isolated
part of a larger, circular supernova remnant.
It was shown by Barlow et al (1994) that the IS and OS of P~Cygni
are most likely shock--excited  
for their expansion velocities
are sufficiently high and there are insufficient Lyman photons
from the star to sustain the level of ionization. In which case,
and within the assumption
that the GL in Fig.~1 originated from P~Cygni, shock excitation must also
prevail: this could be  confirmed  by a tentative  association 
of the eastern arcs of the GL with diffuse (ROSAT All Sky  Survey) 
X--ray emission
though very much longer X-ray integrations than the 750 s employed in this
survey
are now needed to be certain of a detection. The optical line
emission spectrum of the GL may though be that of shocked ambient
gas that existed prior to the LBV phase 
rather than that of processed ejected material.

The radial velocities of the newly discovered southern ridge
can be seen in Fig. 2 (GL south ridge) to be distributed
around \vsys\ of P~Cygni.
Moreover, the western complex of the GL connects both kinematically and 
morphologically
with a feature in the OS of P~Cygni that is only 60\arcsec\ from star
and undoubtedly part of the outburst that created the OS
(see GL west marked in Fig. 2). Near to the star
it has a radial velocity difference of +80 \kms\ (equal to that of 
the far side
of the expanding OS) and reaches a difference of 
+130\kms\ 7\arcmin\ to the east of P~Cygni. Significantly, there
are no other similar \oiii\ emitting filamentary structures in the 
rest of  the 89\arcmin\ $\times$ 89\arcmin\ field covered by the
present imagery.  

The discovery of a southern counterpart to this northern arc (Fig. 1)
and the kinematical evidence in Fig. 2 from slit A  suggests that
they could be the edges of some form of outflowing cavity, which greatly
strengthens the association of the GL with P~Cygni: this morphological
and kinematical behaviour is remniscent of an individual lobe
projecting from a bi-polar planetary nebula such as NGC~6302 (though
radiative ionization by the central hot star dominates
collisional ionization by shocks in the lobes of the latter). The behaviour
of the newly discovered western complex of the possible P~Cygni
GL (Fig. 1) is as yet unclear though it appears to again be
receding with a similar radial velocity to that of the eastern
arcs near to P~Cygni (slit B in Fig. 2).

If this is only
a chance superposition of SNR filaments with P~Cygni the positive 
large radial velocity differences (slit B in Fig. 2) could not be explained as 
a consequence of galactic rotation for well separated objects
along the sight line. For the galactic longitude of P~Cygni of
l $\approx$ 75\deg\ the GL would have to be around 
five times further away than P~Cyg (i.e. $\approx$ 9 kpc distant)    
to give such a radial velocity difference by galactic rotation.
In these circumstances interstellar extinction at this galactic latitude of 
b $\approx$ 1.3\deg\ would make the GL unobservable.
A complete map of \oiii\ line profiles over
the whole GL of P~Cygni is now obviously needed to explore further its
possible kinematical association with the star.  

The evidence linking the proposed GL with R~143 is more tenuous:
it is primarily morphological as the bright ridge shown in Fig. 3b 
(and see an HST image of
this in figure 5 of Weis 2003) projecting from the star is the 
starting point for two of the northern fainter nebulous arcs (see
Figs. 3a \& b). Also to stengthen this correllation 
the proposed GL candidate projecting from
R~143 is around 20\arcsec\ in extent to give an apparent linear extent
of $\approx$ 5 pc which is very comparable to the $\approx$ 9 pc
apparent extent of the P~Cygni GL shown in Fig. 1.
Furthermore, there is some kinematical evidence of an
outflow along these ridges close to R~143
(see figures 7 \& 8  in Weis 2003). 

Detailed kinematical investigation of the R~143 GL candidate is 
required to be certain of this origin for there
are many similar filamentary structures in the rest of the halo
of 30~Dor where R~143 resides.

\subsection{Formation of a  GL by an LBV star}

As the balance of evidence seems to favour the creation of a GL 
by P~Cygni  it is 
interesting to explore
the mechanisms by which such a feature could be generated.  
Most likely  these considerations could possibly also apply to 
the similar star R~143 and other LBVs exhibiting similar GL
phenomena. 

Firstly, the suggestion by Meaburn et al (2000) 
that the P~Cygni GL could be 
a collimated stellar outflow trailed by the passage of P~Cygni
through its local insterstellar medium is now completely discounted
by the latest imagery in Fig. 1. This early idea seemed
possible when the GL was 
considered to be only a simple one--sided ridge (e.g. Meaburn et al 2000).

Under the assumption that the GL candidates do originate in 
P~Cygni and possibily  R~143 it is initially informative to 
estimate the 
dynamical ages of the P~Cygni GL compared with those 
of its IS (880 yr) and OS (2400 yr) as given by 
Barlow et al
(1994) and  Meaburn (2000). Take the apparent eastern extent of
GL from  P~Cygni as 12\arcmin\ (Fig. 1) and at this extremity its radial
velocity difference from P~Cygni as +130 \kms\ (Meaburn et al 1999) then, for
an outflow away from the star, along a line tilted at $\theta$ degrees to the
plane of the sky, a dynamical age of 5 $\times$ 10$^4$ tan~$\theta$ yr is
estimated. For reasonable values of $\theta$ (say $\leq$ 45\deg) 
this age is around
30 times that of the  dynamical age of OS. This would imply
that the GL features in Fig.~1 furthest from the star were formed  well
prior to
the recent eruptions of the star that caused the IS and OS. As the 
GL features nearest to the star appear to be  connected with the OS it would
imply that the generation of the GL occurred right up to the
eruptions that created the OS. 

However,
Humphreys and Davidson (1994) argue that the duration of the LBV phase of
a 50 \msun\ star, after it leaves the main sequence, 
is $\geq$ 2.5 $\times$ 10$^{4}$ yrs  in which case the GL
could simply be a consequence of  sporadic LBV eruptions over 
$\geq$ 5 $\times$ 10$^{4}$ yr. Incidentally, Lamers et al (2001) consider the 
observed dynamical ages
of all of the then known LBV nebulae to be between
1 and 7$\times$10$^{4}$ yr with the exception of that of P~Cygni
which they say is much younger. However, they have only considered
this age for the IS of P~Cygni and not even of
the older OS. The dynamical age of the P~Cygni GL derived here is therefore
well within this observed range of ages for other LBV ejecta
and suggests that P~Cygni is not unusual in this respect.  

Morris et al (1999) discovered a massive equatorial torus in the LBV 
$\eta$~Carinae stellar system (and see Smith et al 2002). 
They suggested that this existed prior to
the star's LBV phase and shaped the ejecta of the subsequent LBV eruptions into
the bi-polar outflows that are now observed. A similar torus has being found
around P~Cygni by Meaburn et al (2000) with its axis parallel to that
of the two eastern ridges of the GL (see Fig. 1). If these do form 
one coherent `lobe' ejected from P~Cygni its shaping by the central torus 
also seems possible. The western parts of the P~Cygni GL do not fit
easily into such a model: they are similarly receding from the observer
and erratically distributed. 
In fact the whole of the P~Cygni GL could delineate the shocked walls 
of an irregular cavity shaped by this torus and
formed by the energetic wind of the star when on the main sequence.

Two anomalies require some consideration i.e. the star P~Cygni is
to the south of most of the proposed GL shown in Fig. 1 and
most of the GL nebulosity is receding from the star (Fig. 2). 
 An explanation may be
found in the motion of the star with respect to its local medium
in the plane of the Galaxy (calculated incorrectly in Meaburn et al 2000).
For this purpose the measured proper motion, PM,
of P~Cygni (Hipparcos PM(RA) = -3.53 $\pm$ 0.39 mas yr$^{-1}$ 
and 
PM(DEC) = -6.88 $\pm$  0.42 mas yr$^{-1}$) can be combined with 
the stellar radial velocity (arrowed in Fig. 2).   
For the stellar position $l$ = 75.87\deg\ and $b$ = 1.311\deg\
and Oort's constants A = 14 \& B = -12 \kms\ kpc$^{-1}$ then only
a residual southerly PM(DEC) = -3.52   $\pm$  0.42 mas yr$^{-1}$ 
remains  with respect to the P~Cygni local medium. The 
residual PM(RA) is, within the errors, equal to the measured Hipparcos value.
Over the dynamical age of the P~Cygni GL this residual PM(DEC) would amount
to a southerly displacement of around 2.9\arcmin\ which goes
some way to explain the observed angular displacement of the star.
For a distance of 1.8 kpc then a tangential velocity V$_{tan}$ = 29 \kms\
is indicated, again with respect to the local medium of P~Cygni. 

At 1.8 kpc distance the medium local to P~Cygni
will have \vlsr\ = 12 \kms\ whereas the
observed systemic \vlsr\ = -8 \kms\ (arrowed in Fig. 2 and using 
\vlsr\ - \vhel\ = 17.5 \kms) to give a radial velocity 
difference of -20 \kms\  for the star with respect to its local medium;
which,  when combined with the tangential velocity difference, indicates
that the star is moving through its ambient gas at $\approx$ 35 \kms.
It is likely that any long term ejecta from the star, such as that
required to form the GL, will have been subjected to a substantial 
`breeze' that
could have resulted in an additional, receding, radial velocity component 
being imparted to the GL, as
observed.

\section{Conclusions}

The GL apparantly projecting from P~Cygni has now been shown
to have a southern counterpart on the eastern side of P~Cygni. Also
a more complex counterpart has been discovered to the west of the star.

The overall apparent extent of this GL is now found to be 9 pc.

These structures emit the \oiii\ line strongly enhancing their contrast
against the confusing emission from the ambient ionized gas along the
same sight--lines.

These newly discovered structures are now shown to have been detected
kinematically in previous spectral observtions. Although receding
radial velocities dominate, kinematical association with P~Cygni is 
strengthened, for the newly discovered southern ridge of the GL,
on the eastern side of P~Cygni, has  radial velocities 
on either side of \vsys\ of this star. 

It is proposed that the P~Cygni GL phenomenon was formed by continual
activity between the age of the OS (2400 yr) and the dynamical
age ($\approx$ 5~x~10$^{4}$ yr) 
of the extreme extent of the GL. Within this interpretation
sporadic LBV eruptions
over this extended period appear to have ocurred. 

The irregular shape of the GL could then be a consequence
of the shape of the cavity formed by the wind of the 50~\msun\  
star when on the main
sequence immediately prior to its LBV phase. 

Similar GL--like features, with an apparent extent of 5 pc, 
are shown to project  from the LMC, LBV star R~143.
Their presence strenghtens the possibility that these and their
P~Cygni equivelents are associated intimately with these stars and 
not just chance alignments of un-related supernova filaments some of 
whose characteristics they share.

\begin{acknowledgements}

We acknowledge the excellent support of the staff at the Skinakas,
San Pedro Martir and La Silla (obtained for us in the Service Observing
programme by Emanuela Pompei)
observatories during these observations. JAL gratefully acknowledges
financial support from CONACYT (M\'ex) grants 32214-E and 37214 and
DGAPA-UNAM IN114199. Skinakas Observatory is a collaborative project
of the University of Crete, the Foundation for Research and
Technology-Hellas, and the Max-Planck-Institut f\"{u}r
extraterrestrische Physik. MPR is supported by the IRCSET, Ireland.

\end{acknowledgements}

\end{document}